\documentclass{jaa}
\usepackage[authoryear]{natbib}
\usepackage{adjustbox}

\usepackage{graphicx}

\begin{document}\sloppy

%%paper title
%%For line breaks \\ can be used within title
\title{Investigating the properties of nearby young moving groups\\ using \textit{Gaia}DR3}

%%author names are separated by comma (,)
%%use \and before the last author name
%%use a * along with the number separated by comma
%% for the  author for correspondence
%%\textsuperscript{number} is used for affiliation
%%\affilOne, \affilTwo etc., upto \affilTwentyfive is possible
%%Please note the first letter after \affil is capitalised in the command
%%

\author{M. Lad\textsuperscript{1,*}, K. Ujjwal\textsuperscript{2}, Blesson Mathew\textsuperscript{2}, S. Raksha\textsuperscript{2}, R. Arun\textsuperscript{3}, Sreeja S. Kartha\textsuperscript{2} and V. Valsan\textsuperscript{2}}
\affilOne{\textsuperscript{1}Departamento de Física, Facultad de Ciencias Exactas, Universidad Andrés Bello, Av. Fernandez Concha 700, Las Condes, Santiago, Chile\\}
\affilTwo{\textsuperscript{2}Department of Physics and Electronics, CHRIST (Deemed to be University), Bangalore 560029, India\\}
\affilThree{\textsuperscript{3}Indian Institute of Astrophysics, Sarjapur Road, Koramangala, Bangalore 560034, India}

%%escape two column mode for title, affiliation and abstract
%%by giving \twocolumn command as shown

\twocolumn[{

\maketitle

%%include \corres to print the corresponding author Email id
\corres{m.lad@uandresbello.edu}

%%include \msinfo for
%%manuscript information such as
%%received, revised and accepted dates
%%
\msinfo{18 July 2025}{13 August 2026}

%%abstract
\begin{abstract}
Moving groups, which are gravitationally unbound collections of stars spread over large portions of the sky, pose challenges to their identification. Analyzing the three-dimensional spatial motion of stars is one method to identify their members. Despite several studies in the past, a reliable and conclusive catalog of moving group stars is currently lacking. Our objective is to present a most recent and updated catalog of nearby young moving group (NYMG) candidates in the solar neighborhood and investigate their properties using updated science data from the \textit{Gaia} Data Release 3 (DR3). We searched for candidates of twelve NYMGs within a distance of 150 pc using \textit{Gaia} DR3. To determine the membership, we employed the Bayesian algorithm BANYAN-$\Sigma$. We compiled a total list of 3,153 NYMG candidates, of which 1,651 were new candidates. We also assessed the credibility of the literature defined 'Good-box' criterion with the latest comprehensive catalog of NYMGs from our study. We homogeneously estimated the ages of the NYMG candidates using \textit{Gaia} DR3 photometry data. The analysis revealed that our NYMG candidates display a large scatter on the CMDs resulting in a large range of estimated ages from isochrones. Additionally, we conducted an infrared (IR) excess analysis to identify disk candidates among our final sample. Our spectral energy distribution (SED) analysis found 51 stars with IR excess. We present a largely inclusive list of all the NYMG candidates within 150 pc of Solar neighbourhood using \textit{Gaia} DR3. The wide range of ages obtained from isochrone fitting underscores the need for more robust age analysis techniques to accurately determine the ages of NYMG members. The presence of IR excess in the 51 stars confirms the existence of disks indicating that they could be potential candidates for exoplanet detections.
\end{abstract}
%%insert keywords separated by 3 hyphens using \keywords{words}

}]
%%close the twocolumn escape here

%%include \doinum{number}for the DOI number in the header
%%include \volnum{number} for the volume number in the header
%%include \year{yyyy} for  year of publication in the header
%%include \pgrange{num--num} page range of article in the header
%%include \artcitid{num} for the article citation id
%%include \lp to print last page of the article
%%include \setcounter{page}{pagenum} for the exact starting page of the article

\doinum{12.3456/s78910-011-012-3}
\artcitid{\#\#\#}
\volnum{000}
\year{0000}
\pgrange{1--}
\setcounter{page}{1}
\lp{1}

\section{Introduction}
\label{sect:intro}
Moving groups are groups of stars in the sky possessing similar space motion and age, but spread over a wide portion of the sky. \citet{ 1869RSPS...18..169P} and \citet{kapteyn1905reports} found that a few stars exhibited similar space velocities to the rest of the (field) Galactic stars. This led to the discovery of co-evolving groups of stars, sharing similar kinematics, known as Moving Groups. There are several hypotheses proposed, to explain the formation mechanisms of moving groups. One hypothesis, for the formation of moving groups, is the cluster disruption mechanism. A likely scenario for cluster disruption could be the interaction of open clusters with massive objects in the Milky Way, such as giant molecular clouds, which stretches out the cluster components into a `tube-like' structure \citep[eg.][]{jeans1922motions, bok1934stability,  eggen1965some}. Based on the cluster disruption hypothesis, \citet{eggen1965some} suggested that the moving groups can be considered as the \textit{missing link} between the open clusters/associations and the field stars. Another hypothesis considers moving groups as a result of the resonant dynamical structures in the galaxy. \citet{famaey2007hyades} studied the Hyades association and found that it is a mixture of stars evaporated from the Hyades cluster and some old field stars trapped at the resonance by a spiral perturbation while forming the stream. However, studies such as ~\citet{tang2019discovery}, \citet{2019A&A...627A...4R}, \citet{2020A&A...638A...9R} and \citet{gagne2021number} have reported disrupted tidal streams around the open clusters using high-quality data from \textit{Gaia} Data Release 2 (\textit{Gaia} DR2 hereafter), suggesting cluster disruption as a most favorable scenario for the origin of the moving groups.
\par

Identifying the candidates of a moving group based on their position is difficult since they are spread across several cubic parsecs in the sky. The nearby young moving groups (NYMGs hereafter) are the most immediate dissipation product of the younger associations in the solar neighborhood \citep{lamers2005analytical, gieles2006star, gieles2007effect}. Spectroscopic analysis of such moving group stars \citep{miret2020dynamical, 2010A&A...521A..12M, zuckerman2004young} show that they are mostly young, with their age ranging from 8 to 100 Myr (with the exceptions of older populations with ages $>$ 400 Myr, as for example, Coma Berenices \citep[700-800 Myr;][]{tang2019discovery} and Group X \citep[400 Myr;][]{tang2019discovery}). The young nature indicates that candidates of NYMGs have not had enough time to disperse their galactic/space velocities significantly from their host association.
As a result, the stars of NYMGs possess common space velocities with typical velocity dispersion of $\approx$ 3 km~s$^{-1}$. Therefore, analyzing the six-dimensional (6D) kinematic properties (3 positional coordinates (XYZ) and 3 space velocity coordinates (UVW)) of each NYMG candidate becomes crucial for identifying their memberships. Furthermore, on analyzing the velocity distributions of the moving group stars, \citet{zuckerman2004young} found that they display an overdensity, indicating that the space velocities of moving group stars are within a specific limit. They termed this limit on velocities as the ``Good-box" criterion. Until recently, the trigonometric parallaxes and other kinematic properties were only available for the massive coeval co-moving stars, which comprise a very small portion of NYMGs. Several methods and algorithms have been put forth to determine the membership of the low-mass NYMG candidates with missing kinematic information, for e.g., convergent point method \citep{eggen1958stellar,galli2012new}, various photometric and/or space velocity selection criteria \citep[][]{higashio2022disks, zuckerman2004young, kraus2014stellar, riedel2017lacewing}, iterative model containing 3D ellipsoidal models \citep{2019MNRAS.486.3434L} and analysis using bayesian statistics \citep{gagne2014banyan,gagne2018banyana, gagne2020mu}. Although analyzing the membership probabilities of these moving group stars is a challenging task, the availability of precise and high-quality astrometry data aids in the identification of moving group candidates. 

The third data release from \textit{Gaia} mission \citep[hereafter \textit{Gaia} DR3;][]{2023A&A...674A...1G} provides complete 6D astrometry solutions i.e., right ascension ($\alpha$), declination ($\delta$), proper motions- $\mu_{\alpha}$, $\mu_{\delta}$, parallax ($\pi$) for $\approx 1.46 $ billion sources and provides radial velocity ($\rho$) for $>33 $ million sources \citep{katz2022gaia, blomme2022gaia}. The availability of 6D kinematic data from \textit{Gaia} DR3 for the faint low-mass stars and the high accuracy in astrometry, makes analyzing the membership probability of NYMGs more accurate. Moreover, \textit{Gaia} DR3 can also aid in identification of new NYMG candidates that were previously unrecognised due to lack of comprehensive and precise astrometric data. Previous data releases from {\it Gaia} have been extensively used to identify and characterize the moving groups \citep{gagne2018banyanc, ujjwal2020analysis, luhman2022census}. The study by \citet{gagne2018banyanc} was focused on the addition of new candidate members to the existing catalog of moving group stars. On the other hand, \citet{ujjwal2020analysis} reanalyzed the membership probability of NYMG candidates reported in \citet{riedel2017lacewing}, using \textit{Gaia} DR2. Although \citet{ujjwal2020analysis} did not explore the possibility of new detections, they discover that the membership probability changed for several moving group members when analysed with updated \textit{Gaia} astrometric data. Other studies like \citet[][]{2006ApJ...649L.115Z, schlieder2010beta, 2015MNRAS.447.1267M, luhman2022census}, etc., have focused on the analysis of a specific moving group. Even though there are a handful of studies done using the data from \textit{Gaia} releases, it needs to be noted that there is no comprehensive list of NYMG candidates compiled using high-precision data in the \textit{Gaia} era.

In our study, we attempt to provide a fairly comprehensive and updated census of moving group candidates belonging to 12 NYMGs within 150 pc and study their stellar properties using \textit{Gaia} DR3. To identify the NYMG candidates we used a Bayesian model selection tool: Bayesian Analysis for Nearby Young AssociatioNs (BANYAN) $\Sigma$ \citep{gagne2018banyana}, which is one of the widely used tools for determining membership probability of nearby young associations in the solar neighborhood. \citet{ujjwal2020analysis} in their study put forth open questions regarding the need to redefine the Good-box criterion, which was first defined in \citet{zuckerman2004young}. We analyzed the space velocity distribution of all the candidates and verified the applicability of the Good-Box criterion to our sample. Based on our analysis we proposed a refined Good-box on the space velocity plots, which can be considered as new constraints on space velocities for identifying the NYMG candidates. Some studies in the past \citep[e.g.][]{mamajek2014age, malo2014banyan, 2016MNRAS.461..794P, ujjwal2020analysis} have reported a significantly widespread in the age of the moving groups, but with an incomplete set of samples. In this work, we evaluated this problem from a re-analysis of the ages of moving group candidates in the homogeneous manner.

The fact that NYMG stars are not part of any cluster or association makes them the ideal ensemble for exoplanet studies \citep{chauvin2010deep, biller2013gemini}. Planet formation and evolution take place in the circumstellar disk \citep[eg.,][]{marois2008direct,lagrange2010giant,gilbert2022flares}. Particularly, the presence of evolved circumstellar disks for many of the moving group stars makes them prospective candidates to search for exoplanets \citep{bowler2013planets, bowler2016imaging, nielsen2019gemini, demangeon2021warm}. The flux excess in the infrared (IR excess) is considered a sign of youth and indicates the presence of a circumstellar disk \citep[e.g. ][]{lawler2012debris, espaillat2014observational}. From our updated list of NYMG candidates, we identify stars that show IR excess and thereby evidently host a circumstellar disk. We catalog the disk candidates in the NYMGs to aid future studies on disk properties and dynamics.
 
The paper is organized as follows. The sample used in our study and the membership analysis performed is described in section \ref{sample} The sample chosen for additional investigation, which forms the basis of our main findings in this study, is described in detail. Additionally, a brief discussion of the new bonafide candidates found in this investigation is also included in section \ref{sample} Successive discussions on space velocity analysis and age estimations are presented in section \ref{sp vel}  and section \ref{age}, respectively. Analysis of IR excess candidates is described in section \ref{ir excess}. Major results from this study are summarised in section \ref{discuss}

\section{Sample Selection}

\label{sample}
We retrieved all the sources within a distance of 150 parsecs reported in \textit{Gaia} DR3, having \verb|parallax_over_error| (parallax/parallax\_error) $>3$. This condition was implemented to separate the sources with reasonable astrometry solutions from those with higher uncertainties in parallax and resulted in an initial sample of $\sim$ $2.1$ million sources. Then, we implemented the selection criteria of Re-normalised Unit Weight Error (RUWE) $<$ 1.4, where the single-star model provides an excellent fit to the astrometric observations. Any value of RUWE $>$ 1.4 indicates that the source is non-singular or there is an anomaly associated with its detection \citep{blanchard2020euclid, stassun2021parallax}. Since the proper motion estimates significantly affect the membership analysis, we only selected the stars having $|\mathrm{pmRA/pmRA_{error}}|$ \& $|\mathrm{pmDE/pmDE_{error}}|$ $>3$. After applying all the above-mentioned constraints a sample of $967,932$ stars was obtained and was used for further analysis.

For identifying the membership probability of the stars we employed BANYAN $\Sigma$ algorithm \citep{malo2012bayesian, gagne2018banyana}, which is a next-generation Bayesian inference tool designed for analyzing membership probability of 27 nearby kinematic associations within 150 pc of the solar neighborhood. 
%Later, with the release of \textit{Gaia} DR2, newer candidates were identified using the algorithm \citep{gagne2018banyanc}}. 
It is constructed using multivariate gaussian models in 6D space ($XYZUVW$). The algorithm makes use of the kinematic information (i.e., $\alpha$, $\delta$, $\pi$, $\mu_{\alpha}$, $\mu_{\delta}$ and $\rho$) of all the known bonafide candidates belonging to 27 coeval associations, which were compiled from various surveys in literature and are referred as \textit{literature bonafide} candidates hereafter. 

BANYAN-$\Sigma$ makes use of this \textit{bonafide} candidates as a training set for constructing multivariate Gaussian models for different associations as well as for \textit{field} stars. The multivariate Gaussian models are fitted to the position and velocity distribution of the young associations to define the NYMGs. The ellipses help to determine the star's membership probability to a particular NYMG. Using the six parameters ($\alpha$, $\delta$, $\pi$, $\mu_{\alpha}$, $\mu_{\delta}$, $\rho$) of the stars, the maximum likelihood of them belonging to a certain association is calculated.  The six parameter consideration makes BANYAN-$\Sigma$ less approximated and more accurate than its predecessor BANYAN- II \citep{2014ApJ...783..121G}. BANYAN $\Sigma$ can predict the membership probability even with the absence of full 6D kinematic information of the star by assigning the parameters within the 1$\sigma$ dispersion of the average distance/radial velocity of the entire group. However, the lack of full 6D kinematics for a significant number of candidates will result in a lower recovery rate of their true candidates by the algorithm and a larger number of contaminants from field stars. Taking into account all of the above advantages, BANYAN- $\Sigma$ serves an appropriate choice for the membership analysis in the current study.

Our sample of $967,932$ stars was passed through a membership test in the BANYAN $\Sigma$ algorithm. We found that $3,144$ stars are identified as NYMG candidates. The rest of the stars either belonged to other associations or were identified as `field' stars. BANYAN $\Sigma$ consists of 1,095 candidates of 12 NYMGs, namely 118 Tau (118TAU), AB-Doradus (ABDMG), $\beta$- pictoris (BPMG), Carina (CAR), Carina near (CARN), Columba (COL), $\epsilon$-Chameleontis (EPSC), Octans (OCT), Tucana-Horologium (THA), 32 Orion (THOR), TW-Hyades (TWA), $\chi^1$ FOR (XFOR) which we consider for the present study. Out of 1095 stars, $387$ were \textit{bonafide stars} as stated in \citep[which are basically compiled from literature sources]{gagne2018banyana} and the bonafide members from \citet{gagne2018banyanc} (which were identified using \textit{Gaia} DR2 data). Among these 387 candidates, 230 had \textit{Gaia} DR3 detections with a RUWE of less than 1.4. Notably, 208 of these bonafide candidates were already included in our initial sample of 3,144 stars. The inclusion of the remaining 22 (230 - 208) candidates led to the catalog comprising 3,166 moving group candidates (3,144 + 22).

After an extensive literature review, we identified 13 candidates out of the initial 3166 whose membership in moving groups had been previously rejected in various studies. These rejections were based on a range of criteria, such as spectroscopic investigations \citep{kraus2014stellar, 2016MNRAS.455.3345B}, photometric analyses \citep{2017ApJS..228...18G}, and kinematic selections \citep{2005ApJ...634.1385M}. The removal of these 13 stars resulted in the number of confirmed moving group candidates to be 3,153. We propose that our final catalog of 3,153 candidates can be used as a comprehensive list of known moving group candidates within 150 pc. We posit this catalog to be the most trustable aid to any upcoming studies of kinematic groups in the solar neighborhood. In our catalog, radial velocity estimates from \textit{Gaia} DR3 are available for 1,314 stars whereas photometry in all the \textit{Gaia} pass-bands is available for 1,327 stars.

 Out of the total sample of 3,153 NYMG candidates, 1,502 stars were previously identified as NYMG candidates in prior studies \citep[e.g.][]{gagne2018banyana, gagne2018banyanb, gagne2018banyanc, 2015ApJ...798...73G, 2019AJ....158..122K}. It is important to note that 379 (out of 1502) stars have different membership than the one we obtained in our current study. Consequently, we have identified 1,651 stars (3,153 - 1,502) as new NYMG candidates, marking the first recognition of their NYMG candidacy.  Within this group of 1,651 prospective members, 29 stars exhibit membership probabilities exceeding 90\%. These stars can be confidently considered as new bonafide candidates of moving groups and can be a valuable addition to the existing list of bonafide candidates reported in \citet{gagne2018banyana} and \citet{gagne2018banyanc}.

\section{Results and Discussions}
\label{result}
\subsection{Space velocity analysis of NYMGs and the evaluation of the Good-box criterion}\label{sp vel}
Stars in NYMGs in the Solar neighborhood, due to their proximity, span a large portion of the sky and thus, can be confused with other field stars that are not a part of any coeval group. Therefore, to isolate the coeval stars from the large sample of stellar populations, it is important to obtain $(X, Y, Z)$ and $(U, V, W)$ coordinates of the stars. $(U, V, W)$ coordinates primarily depend upon the astrometry parameters like proper motion and radial velocities of the stars. Prior to the advancements of deep sky surveys like \textit{Gaia} , numerous low-mass stars lacked radial velocity estimates, causing alteration in $U, V, W$ estimates.

The Good-box criterion was introduced by ~\citet{zuckerman2004young}\footnote{Although the idea of Good-box is mentioned in \citet{zuckerman2004young}, an extensive analysis was not performed in this review paper.} in their investigation of nearby moving groups. They observed that majority of their moving group candidates displayed an apparent over-density in space velocity diagrams. A boundary was outlined for this over-density forming a square (``box") region on the space velocity plot (see their Fig. 6). This region was termed as ``Good-box" criterion and it imposed constraints over the velocities of NYMG stars. Its dimensions in terms of $U,V,W$ values were in the range of $0$ to -$15 $ km~s$^{-1}$, -$10$ to -$34 $ km~s$^{-1}$, $3$ to -$20$ km~s$^{-1}$, respectively,  It implies that the nearby stars (within $60\,pc$ of distance), that had space velocity in the range of Good-box dimensions can be considered as an NYMG candidate. For defining this criterion they assumed that the UVW range for all young nearby stars should be within 10 km~s$^{-1}$ of the average $UVW$ of `Eggen's Local Association' \citep[][]{jeffries1995kinematics}. They obtained 6D kinematic parameters from the Hipparcos catalog to calculate $UVW$. While this criterion has been effective in identifying NYMG candidates historically, \citet{ujjwal2020analysis} highlighted the necessity of redefining this criterion in light of the \textit{Gaia} era. In this study, we revisit the conventional Good-box criterion to explore how the inclusion of newly identified NYMG candidates that are based on modern Bayesian membership analyses would affect its kinematic extent in velocity space. By incorporating this more complete sample, we examine how well the original Good-box captures the current population of NYMG members and highlight its limitations in the era of update data.

\begin{figure}
\includegraphics[width= 0.5\textwidth]{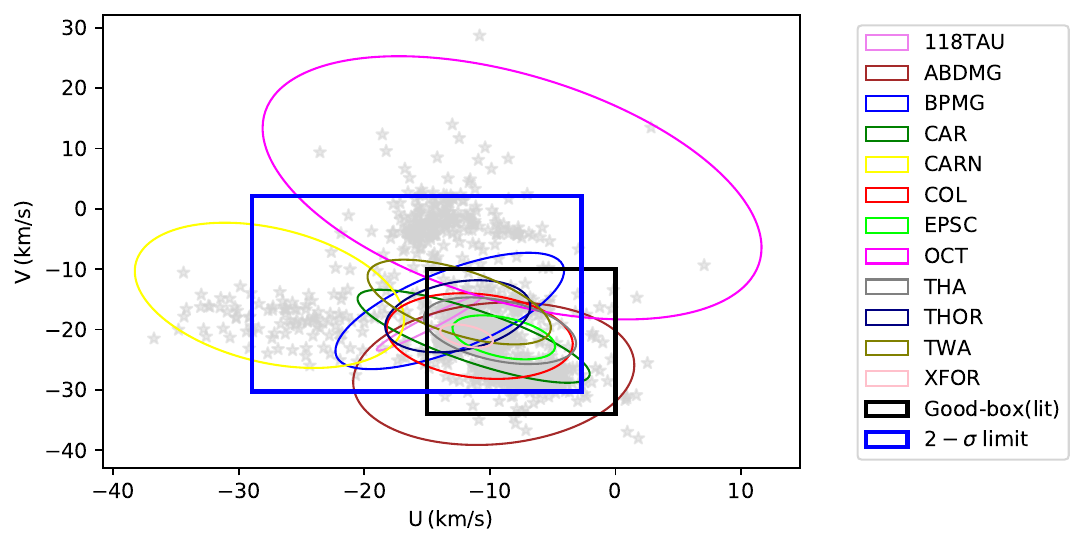}
\includegraphics[width= 0.38\textwidth]{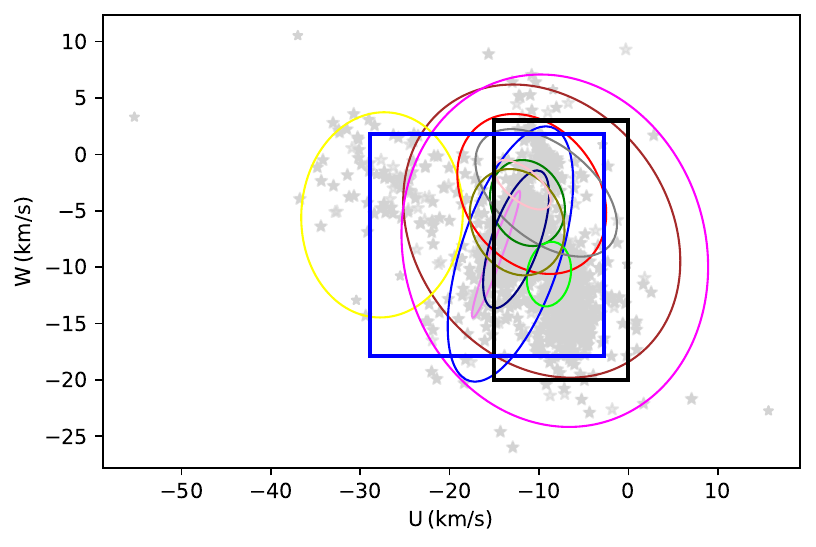}   
\includegraphics[width= 0.38\textwidth]{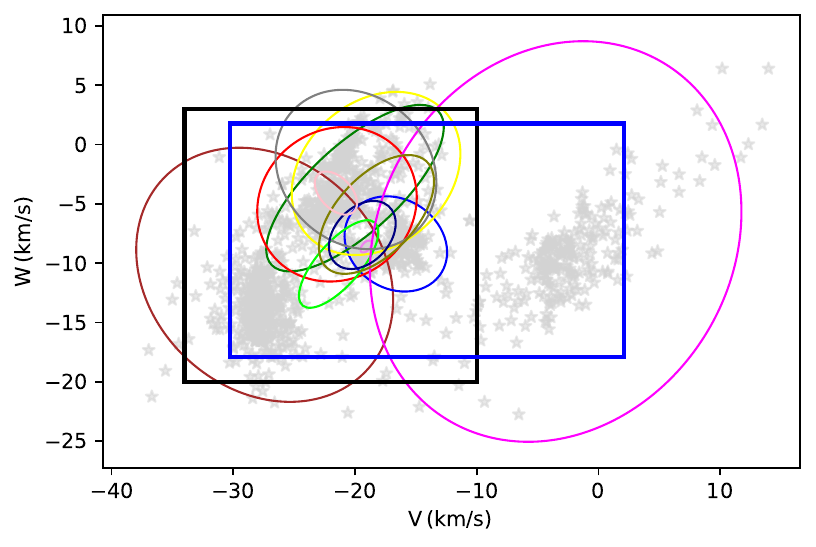}
\caption{Representation of all moving group stars in space velocity plane. Space velocity ellipses for each moving group are color-coded as given in the label in the first plot. The grey stars represent all the NYMG candidates we identified in our study. Each group occupies a certain patch in the velocity plane, represented by different ellipses. The black rectangle represents the Good-box defined by \citet{zuckerman2004young}. The blue solid line rectangle represents the dispersion criteria of 2$\sigma$ on the Good-box, which is the region we propose to be a refined Good-box based on our results.}
\label{sp gb}
\end{figure}

We calculated the space velocities for our NYMG sample using the matrix equation given in \citet{johnson1987calculating}. This equation derives $UVW$ values using astrometric parameters such as $\alpha, \delta$, $\mu_{\alpha}$, $\mu_{\delta}$, $\rho$, $\pi$, which we used from \textit{Gaia} DR3. After calculating the values of $U, V, W$ for each NYMG, we plotted their velocity ellipses in $\mathrm{U\,v/s\,V}$, $\mathrm{U\,v/s\,W}$ and $\mathrm{V\,v/s\,W}$ planes, as shown in Figure \ref{sp gb}. The grey stars in Figure \ref{sp gb} represent all the NYMG candidates identified from our study. The over-density of spatial velocities is observed for stars belonging to the same NYMGs. The velocity ellipses for different moving groups are plotted with different colors. From Figure \ref{sp gb}, we can conclude that the old Good-box (black solid line) does not encompass all the candidates from our sample. Hence, this necessitates the need to address and refine the Good-box criterion with the newer release of \textit{Gaia} data. The possible reason that not all stars fit into the Good-box might be due to the lack of kinematic data available for numerous low-mass stars when \citet{zuckerman2004young} defined the criterion. Due to the advent of \textit{Gaia} DR3, we now have accurate kinematic information available for low-mass, low-luminosity candidates.

In-order to refine the `Good-box' criterion, we defined dispersion ($\sigma$) for each of the space velocity components U, V and W. We applied $1\sigma,\,2\sigma\,\&\,3\sigma$ limits on space velocities of our entire sample of NYMG. On visual inspection, we observe that the region obtained from $2\sigma$ dispersion criterion fits decently (in all 3 components) compared to $1\sigma$ and $3\sigma$ (see \ref{1appen}), with $1\sigma$ (Fig. \ref{1sigmaappen}) being too small to include all the groups and $3\sigma$ (Fig. \ref{3sigmaappen}) having a large amount of empty spaces between the groups. The solid blue box in Figure \ref{sp gb} represents the region with $2\sigma$ velocity dispersion, which comprises 95.4\% of our sample. The dimensions of the proposed new Good-box for U,V,W components are -29 to -3 km~s$^{-1}$, -30 to 2 km~s$^{-1}$, -18 to 2 km~s$^{-1}$, respectively. Additionally, to examine how many additional candidates could be identified with our refined Good-box in the future, we looked for the number of stars from entire \textit{Gaia} DR3 sample within 150 pc, possessing space velocities within the new Good-box range. The analysis is detailed in the \ref{2appen}. Our findings indicate that 53,394 stars could be potential candidates belonging to NYMGs (Fig. \ref{app_150pc}), provided that we do not impose very strict constraints over the uncertainties of the astrometry parameters. Although, in recent times, various membership algorithms have been developed for identifying moving group candidates, radial velocity estimates are still unavailable for a significant population of stars. The redefined dimensions of Good-box obtained from the current study shall be beneficial for the stars that are still lacking proper radial velocity measurements. Moreover, iterating the radial velocity of a star over a range of values can be another beneficial approach. Moreover, the redefined constraints of Good-box given in this study, would work as a preliminary test for isolating potential moving group candidates from the stellar populations.

To assess the effectiveness of the redefined Good-box criterion, we evaluate its contamination and completeness within a \boldmath{$2\sigma$} velocity boundary. For this purpose, we examine the number of stars recovered within the updated Good-box from the full Gaia DR3 sample of stars within 150 pc (see Appendix \ref{2appen} for details). In total, 54,522 stars in Gaia DR3 fall within the \boldmath{$2\sigma$} Good-box in velocity space, of which 1,663 are identified as NYMG members in our study.

The contamination rate is given by
\begin{equation}
\frac{N_{\mathrm{false}}}{N_{\mathrm{total}}} = \frac{53,394}{54,522} \approx 97\%,
\end{equation}
and the completeness is
\begin{equation}
\frac{N_{\mathrm{recovered}}}{N_{\mathrm{expected}}} = \frac{1,663}{3,153} \approx 36\%.
\end{equation}

These results demonstrate that, the expanded Good-box captures a fraction of known NYMG members, along with including a substantial number of field stars, leading to high contamination and moderate completeness.
Applying the \boldmath{$3\sigma$} dispersion for the conventional Good-box instead of \boldmath{2$\sigma$} we would recover more candidates, but it would increase the contamination rate significantly. While, applying \boldmath{$1sigma$} dispersion we shall reduce the contamination at the cost of recoveering lesser candidates compared to \boldmath{2$\sigma$}. This highlights the limitations of the Good-box as a standalone selection method and reinforces the importance of probabilistic approaches, such as Bayesian tools, for reliable NYMG identification.

\subsection{Age analysis}\label{age}

Recently, a few interesting studies like \citet{mamajek2014age, 2016MNRAS.461..794P, ujjwal2020analysis} have reported a large scatter in the distribution of stars in their Colour Magnitude Diagrams (CMDs) of NYMGs which could be potentially misinterpreted as an age gradient in the NYMG. The presence of the large dispersion of stars in the CMD could be due to various factors such as unresolved binaries \citep{2017A&A...608A.148Z}, erroneous parallax and photometry parameters \citep{slesnick2008large}, contamination in the sample used for study \citep{hartmann2001age}, spread in rotation velocities of stars \citep[e.g.,][]{niederhofer2015apparent} and limitations in the isochrone models\citep[as shown in][]{malo2014banyan, lee2024revisiting}. Stellar evolution models are subject to a wide range of uncertainties, from the treatment of convection and overshooting, to rotation and magnetic fields. Rotation can induce moderate changes in the observed luminosity, color and brightness. Moreover, it is also shown that rotational mixing can supply fresh hydrogen to the core of the star thereby broadening the main sequence lifetime, mimicking the age spread effect in NYMG \citep{2000A&A...361..101M, 2012A&A...537A.146E, brandt2015bayesian}. The magnetic activities in star can result in inflated radii and cooler temperatures than models may imply. For example, \citet{malo2014banyan} showed that the age of BPMG stars can increase from 5 to 15 Myr by using magnetic evolutionary models, as compared to non-magnetic ones (for $T_{eff} < $ 3500). This can cause discrepency in the derived ages from the isochrones \citep[see][and references within]{feiden2012self}. As a solution to the wide age spread problem observed from isochrones, \citet{bell2015isochronal} provided an alternative technique of self-consistent isochronal age determination method. However, it should be noted that their input sample suffered from contamination from non-moving group stars (which is usually an issue while constructing the isochronal age-dating models) which can in-turn interfere with the age calibration using isochrone. Although, isochrone fitting is subject to high model dependencies \citep[]{bell2012pre,bell2013pre,bell2014pre}, a comprehensive sample of the NYMG members in our study, can aid to more accurate age determination. Moreover, the use of homogeneous data from a single facility like \textit{Gaia} , reduces the scope of uncertainty that arises when observational data is combined from various surveys. \citet{ujjwal2020analysis} demonstrated the usefulness of precise data from \textit{Gaia} DR2 in minimizing the error associated with the age calibration in moving groups. In the current study, we intend to utilize a large sample of NYMG candidates to estimate the ages using photometric data from Gaia DR3

The recent release of \textit{Gaia} DR3 \citep{2023A&A...674A...1G} presents an invaluable opportunity to enhance the precision of astrometric and photometric parameter constraints in comparison to its prior data releases. \textit{Gaia} DR3 exhibits marked improvements, including reduced uncertainties in key measurements such as (1) proper motion \citep[0.02 -- 0.5 mas/yr;][]{2023A&A...674A...1G}; (2) parallax \citep[0.02 -- 0.5 mas;][]{lindegren2021gaia}; (3) position \citep[0.45 mas;][]{2023A&A...674A...1G}. Additionally, \textit{Gaia} DR3 has successfully resolved sources with magnitudes up to 21 in the G band ($G_{\text{mag}}$) for over 1.8 billion sources and provides apparent magnitudes in the red photometer (RP) denoted as $G_{RP}$ for 1.54 billion sources. Notably, the uncertainties associated with G-band magnitudes in \textit{Gaia} DR3 have been reduced to 0.3 milli-magnitudes (mmag) in comparison to the 2 mmag uncertainties in \textit{Gaia} DR2. It should be noted that the errors calibrated in this band are dominated by photometric zero-point technique and are therefore larger compared to most values in the \textit{Gaia} catalog. However, leveraging the best available and higher-precision \textit{Gaia} DR3 data, and adhering to the selection criteria outlined in Section 2, alongside the use of advanced tools such as BANYAN-$\Sigma$ for membership probability assignment, can lead to significant improvements in addressing the factors influencing the observed scatter in the HR diagram that is discussed above. In this study, we intend to investigate how these aforementioned improvements contribute to the observed age ranges among stars belonging to the NYMGs.

\par
\begin{figure}
    \centering
    \includegraphics[scale=0.6]{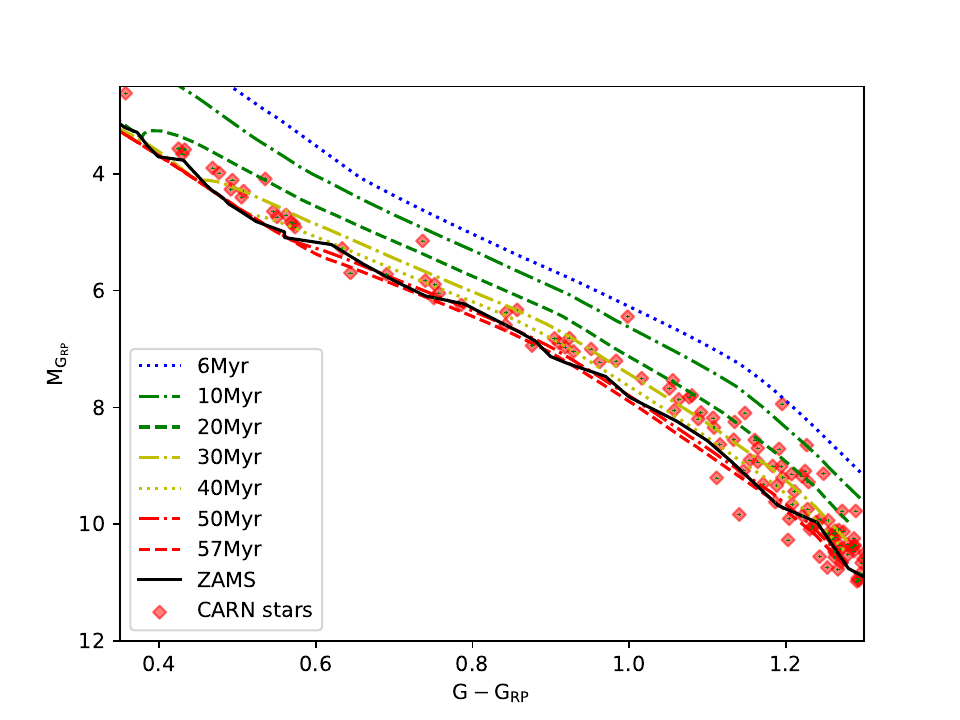}
    \caption{MIST isochrones of different ages(see legend) over-plotted over \textit{Gaia} CMD of CARN stars (\textit{red diamonds}). Zero Age Main Sequence (ZAMS) is represented in \textit{black solid line} and the errors associated with age are given as \textit{green crosses}. As also discussed in the text the age of CARN ranges from $\approx$ 6-57 Myr.}
    \label{isochrone}
\end{figure}

We make use of the absolute magnitude of Gaia's RP passband ($M_{G_{RP}}$), which was obtained using the distances calculated from \textit{Gaia} parallax estimates ($Distance (pc)= 1000 / \pi (mas)$), and the color $(G-G_{RP})$ for plotting CMDs for each NYMG separately. The Gaia DR3 parallaxes were corrected for the parallax zero-point offset (PZPO) using the \verb|gaiadr3_zeropoint| Python package developed by \citet{lindegren2021gaiap}. The resulting PZPO value was subtracted from the Gaia parallaxes to obtain the absolute magnitudes for our candidates.
To estimate the ages we made use of isochrones and evolutionary tracks from `Modules for Experiment in Stellar Astrophysics (MESA) Isochrones and Stellar Tracks' \citep [MIST; ][]{2011ApJS..192....3P, 2013ApJS..208....4P, 2015ApJS..220...15P, 2016ApJS..222....8D, 2016ApJ...823..102C}. Ages for the individual groups were determined by over-plotting the MIST isochrones on the color–magnitude diagrams (CMDs) of each NYMG. To avoid subjective bias, we carried out this procedure using a python-based algorithm that assigns a best-fitting isochronal age to each NYMG candidate. The code takes as input our catalog of candidates containing the values of absolute magnitude and colours for individual candidates, together with a grid of MIST isochrones that provides the corresponding magnitudes and ages. Using these inputs, the algorithm interpolates within the isochrone grid to estimate the age of each candidate based on its position in magnitude–color space. To ensure reliable age estimates, we restricted our analysis to candidates with high membership probabilities (greater than 77\%) and to previously confirmed NYMG members compiled from the literature. This constraint minimizes potential errors in the age estimation arising from low-probability members. Out of the total sample of 3,153 NYMG candidates, a subset of 1,562 high-probability members — including both literature-confirmed and newly identified candidates — was used for the final age determination. Figure \ref{isochrone} shows an example of different isochrones over-plotted on the CMD of CARN candidates from our sample.
From the figure, we observe that there is a large scatter observed for CARN sequence where our candidates match a wide range of MIST isochrones.

Similar to CARN, we successfully estimated isochrone ages for 872 candidates from 12 NYMGs. For our candidates that either lied on/below Zero Age Main Sequence \citep[ZAMS; taken from][]{2013ApJS..208....9P} and below the low mass limit of the MIST isochrones we assign ages as lower limit (i.e., $> 64$ Myr). In this region of the CMD, isochrones corresponding to different ages significantly overlap, leading to strong age degeneracy and consequently large uncertainties in the inferred ages. To avoid assigning unreliable absolute ages in such cases, we report these sources with a lower age bound rather than a specific value. While precise age determination is not feasible for these stars, their positions in the CMD indicate that they are older than $\approx64$ Myr. Thus, we can use them for constraining the overall age range of the NYMG population. We have not considered an extinction factor in the observed \textit{Gaia} magnitudes, because most of the NYMG stars are nearby stars whose magnitudes are not considerably affected by extinction. We obtained the mean ages of each of the groups and compared them with the ages reported in the previous studies, as shown in Table \ref{tab:ages}. ABDMG and CARN exhibit some differences between the ages reported in the literature and those derived in this work. In particular, our estimated ages are younger than the commonly adopted literature values of $150$ Myr for ABDMG \citep{bell2015isochronal} and $150$ - $300$ Myr for CARN \citep{2006ApJ...649L.115Z, 2008hsf2.book..757T}. This difference is likely influenced by the treatment of stars located near or below the ZAMS, for which reliable isochronal ages could not be assigned due to degeneracy. Consequently, this affects the derived mean ages, making them slightly conservative, particularly for the older populations.

To estimate the uncertainties in the derived stellar ages, we first compute the errors in absolute magnitude and color for each star. We calculate the uncertainity in the $M_{RP}$ and $G-Rp$ colour using the standard error propagation formula:

\begin{equation}
    \sigma_{M_{RP}} = \sqrt{\sigma^{2}_{m_{RP}} +\frac{5}{ln 10} \frac{\sigma _{\pi}}{\pi}}
\end{equation}

where $\sigma_{m_{RP}}$ is the error in apparent $Rp$ magnitude, and $\pi$ and $\sigma _{\pi}$ represents the parallax and its uncertainty respectively. We include the values of the uncertainities of colour and absolute values in our estimated ages to generate lower and upper limits on our ages.

\begin{figure}
    \centering
    \includegraphics[width= 1.1\linewidth]{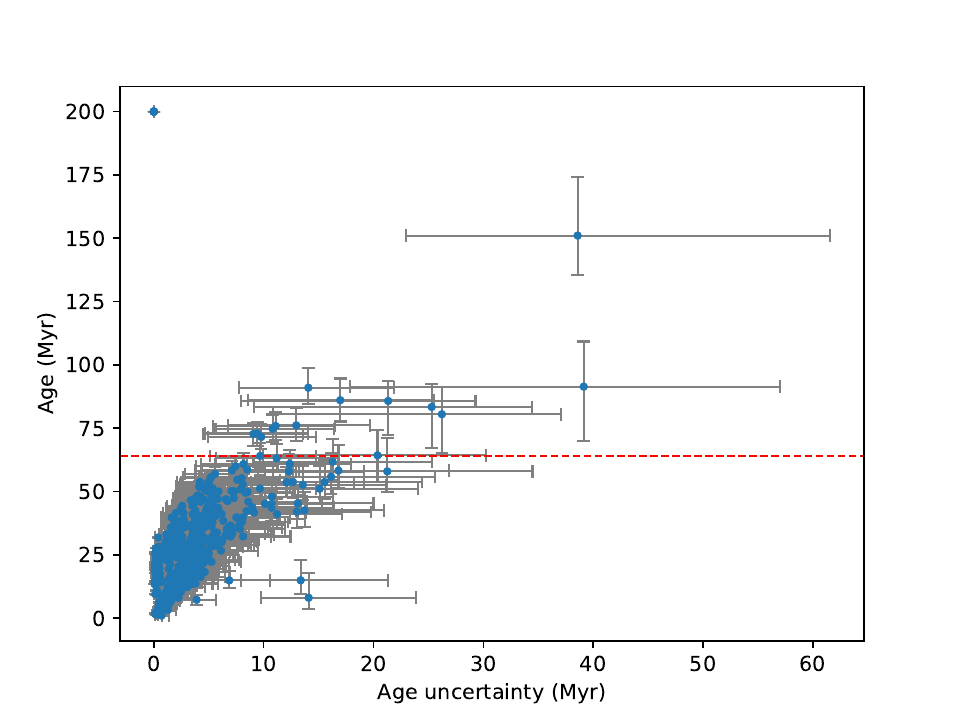}
    \caption{Upper and lower bounds (gray lines) of the ages for individual stars(blue dots). The red dotted line represents ZAMS line coinciding at 64 Myr. We note the increase in uncertainty in the estimated ages as we move towards older stars.}
    \label{fig:age errors}
\end{figure}

Figure \ref{fig:age errors} illustrates the uncertainties associated with the isochronal age estimates for stars in our sample. We observe a clear trend in which the uncertainties increase toward older ages, particularly for stars with ages \boldmath{$>64$} Myr. These stars, which lie near or below the ZAMS(indicated by red dotted line), exhibit significantly larger uncertainties compared to younger sources located above the ZAMS. This behavior reinforces our approach of reporting lower age limits for such stars rather than assigning precise absolute ages, thereby avoiding potentially unreliable estimates.

We observe that the ages of all NYMGs have a large scatter on the CMDs, overlapping on a wide range of isochrones ages. Thus, we computed the mean ages for our candidates that are listed in Table \ref{tab:ages}. Wide ranges in NYMG candidates were also observed in the previous studies from \citet{malo2014banyan, mamajek2014age, 2016MNRAS.461..794P, ujjwal2020analysis}. The large scatter of candidates on the CMD can be confused for a presence of potential age spread. In the current study, we have employed a carefully curated list of known NYMG candidates (with highly probable members) and utilized high-quality upgraded photometry information from \textit{Gaia} DR3 to estimate the ages. Moreover, we believe that taking into consideration several factors such as magnetic stellar evolutionary models \citep{malo2014banyan} and the stellar rotation models \citep{niederhofer2015apparent} could be useful to improve our age estimates in the future. We also believe that further constraints could be achieved on the ages using assertive methods like LDB, although it is beyond the scope of current study.

%From Table \ref{tab:ages}, we note the members of each NYMG match with a varied range of isochrones, which can be confused as a presence of an age gradient within NYMG stars \citep[such wide ranges were also observed in the studies from ][]{malo2014banyan, mamajek2014age, 2016MNRAS.461..794P, ujjwal2020analysis}. In the current study, we have employed a carefully curated list of known NYMG candidates and utilized high-quality upgraded photometry information from \textit{Gaia} DR3 to estimate the ages. Moreover, we believe that taking into consideration several factors such as magnetic stellar evolutionary models \citep{malo2014banyan} and the stellar rotation models \citep{niederhofer2015apparent} could be useful to improve our age estimates in the future. We also believe that further constraints could be achieved on the ages using assertive methods like LDB, although it is beyond the scope of current study. 

\begin{table}
\caption{NYMGs ages compiled from various literature using different age determination techniques and mean isochronal ages compiled from our study.}
\begin{tabular}{|c|c|c|}
\hline
NYMG   &  Mean age  & Literature range    \\
       & (Myr)    & (Myr) \\
       \hline
118TAU  & {\boldmath{$8.40$}}  &   $~10$ \\ \hline 
ABDMG   & {\boldmath{$30.82$}}  &  $100-150\,^{d}$ \\
&&$5-70\,^{d}$ \\
&& $149 ^{+51\,a}_{-19}$  \\ 
\hline
BPMG     & {\boldmath{$21.91$}}  & $20-26^c$ \\
&& $24\pm3^a$ \\ \hline
CAR     & {\boldmath{$11.71$}}  &  $20-50^{c,d}$ \\ 
&&$45^{+11\,a}_{-7}$ \\ \hline
CARN    & {\boldmath{$36.66$}}  & $150-300^b$      \\  \hline
COL    & {\boldmath{$18.79$}} &   $30-50^{b,d}$ \\ 
&& $42^{+6\,a}_{-4}$  \\ \hline
EPSC    & {\boldmath{$11.63$}}  & $1-5^c$  \\ \hline
OCT  & {\boldmath{$28.78$}}  & $30-40^c$    \\  \hline
THA  & {\boldmath{$15.87$}}  &  $30-50^{c,d}$ \\
&& $45\pm4^a$  \\ \hline
THOR    & {\boldmath{$21.03$}}  & $20-30^{c,d}$  \\
&& $22^{+4\,a}_{-3}$\\ \hline
TWA     & {\boldmath{$9.22$ }} & $8-15^a$ \\ \hline
XFOR    & {\boldmath{$16.72$}}  & $25-45^d$ \\ \hline
\end{tabular}
\newline

\textbf{Note}: a- semi empirical isochronal fitting ~\citep{2015MNRAS.454..593B}, b- Li I 6707 \AA {} equivalent width ~\citep{2006ApJ...649L.115Z, 2008hsf2.book..757T}, c- Lithium Depletion Boundary (LDB) technique ~\citep{kraus2014stellar, 2015MNRAS.447.1267M, binks2013lithium, malo2014banyan, shkolnik2017all}, d- isochronal age fitting method ~\citep{ujjwal2020analysis,galli2021chi1, booth2021age, schneider2019acronym, kraus2014stellar, 2008hsf2.book..757T}. Age for 118 TAU is used from ~\citet{mamajek2016}.   
\label{tab:ages}
\end{table}

\subsection{Infrared excess candidates}
\label{ir excess}

Dust particles in the circumstellar disk are heated by incoming starlight and are emitted at longer wavelengths. Thus, an excess of IR indicates the presence of a circumstellar disk \citep{lawler2012debris, espaillat2014observational}. Stars in young moving groups and associations often show excess emission at the IR region, indicating the presence of circumstellar disks \citep{zuckerman2011tucana}. IR excess studies also provide targets for exoplanet search \citep{bowler2013planets, bowler2016imaging, nielsen2019gemini, demangeon2021warm}. Since planet formation occurs in circumstellar disks, studying the evolution of disks can provide valuable constraints on planet formation timescales. Many exoplanets are identified in systems with debris disks. For example, $\beta$ Pictoris was found to host two planets \citep{lagrange2010giant}, HR 8799 was known to have four planets associating with them \citep{marois2008direct, su2009debris} and AU Mic is an interesting system with two exoplanets \citep{plavchan2017discovery, plavchan2020planet, gilbert2022flares}.

\begin{figure*}
    \centering
     \includegraphics[scale=0.5]{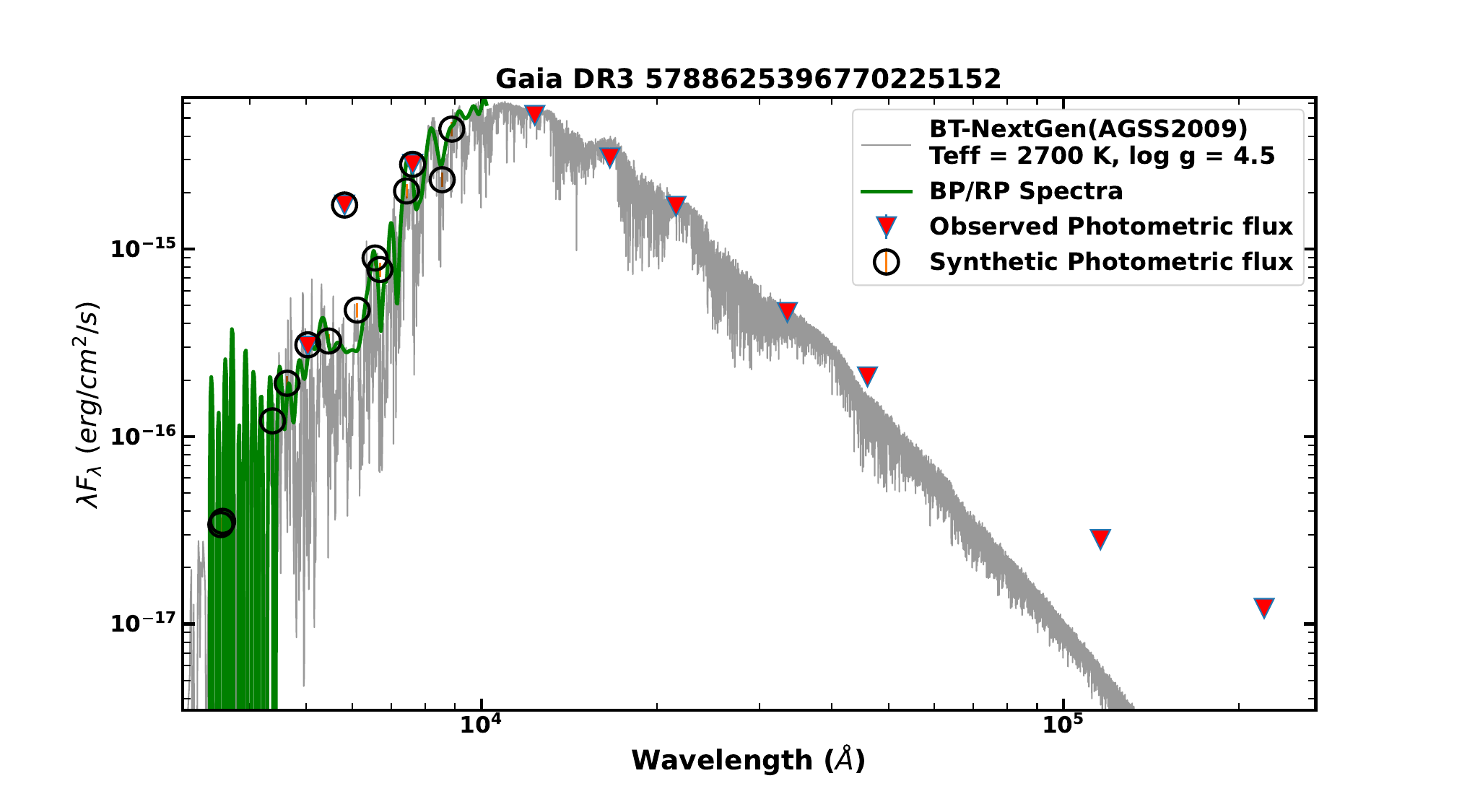}
    \caption{An example of a SED of HD 37852 (\textit{Gaia} 5788625396770225152), displaying an IR excess. The observed data points (red triangles) mark the pre \textit{Gaia} DR3 observed photometric flux. The black hollow circles show the synthetic photometric flux calculated from \textit{Gaia} DR3's BP/RP spectrum. The mean sampled \textit{Gaia} 's BP/RP spectra is shown in green along with BT-NextGen (AGSS2009) spectra model fit represented in gray. The method for SED fitting was discussed in section 2.3 in \citet{shridharan2022emission}}
    \label{fig:sed}
\end{figure*}

NYMGs within 150 pc provide an excellent sample for studying stellar and planetary evolution since the ages span about $\mathrm{10\,to\,100}$ Myr. As our study provides the most updated list of candidates of the 12 moving groups, an investigation of IR excess for this sample could direct us to the candidates with a protoplanetary disk, which can be further analyzed for the study of disk evolution and can be possible candidates for exoplanet systems.

\begin{table*}

\caption{Identified moving group candidates and their estimated stellar parameters using \textit{Gaia} DR3 data. Candidates marked with a star symbol are those with IR excess.The last column represents the status of the candidate, for example, \textit{New}: new candidates, \textit{New b}: new bonafide candidates from this study, \textit{known}: known candidates in literature, \textit{Reclassification}: reclassified candidates in the literature. The NYMG membership of reclassified candidates from this study, should be considered with caution.}
\begin{adjustbox}{width= \textwidth}

\begin{tabular}{|l|l|l|l|l|l|l|l|}
\hline
Gaia DR3 IDs         & Assoc  & Parallax (mas) & U (km/s) & V (km/s) & W (km/s) & Age (Myr) & Status           \\ \hline
5957337926975113472  & BPMG   & 10.06          & -6.8     & -14.67   & -8.41    & 2.2       & New              \\ \hline
3417838118351998976* & 118TAU & 8.95           & --       & --      & --       & 3.81     & New              \\ \hline
3404586976012580992* &118TAU & 9.16           & --      & --       & --       & --      & New              \\ \hline
5166951386298774144* & ABDMG  & 25.82         & -5.22    & -27.94   & -9.72    & 39.15     & Known           \\ \hline
677671008095729152   & TWA   & 22.0         & -9.29    & -29.05   & -13.15  & 17.73     & Reclassification \\ \hline
5949734567882085376  & BPMG  & 9.71           & -5.3    & -16.79   & -8.91    & 6.77      & New b   \\ \hline        
\end{tabular}
\end{adjustbox}
\label{tab: final tab}
\newline 
\newline \textbf{Note:}
      Table \ref{tab: final tab} in its entirety in machine-readable format is attached.
\end{table*}

We use the VizieR database to collect available photometric magnitudes from optical, near-IR, and far-IR wavelengths. We used data from \textit{Gaia} DR3, 2MASS, ALLWISE, Spitzer MIPS (24 $\mu$m band) and Herschel PACS. The search for the disks with ALLWISE is limited by contamination. The point spread function (PSF) for the W4 band of ALLWISE has a full width at half maximum of 12”, allowing the multiple point sources in this radius to blend and produce a false-positive IR excess.  
To avoid contamination and ensure good quality in the photometric measurements, we use the following criteria: ALLWISE contamination and confusion flag -- $cc\_flg =\,'0000'$,  profile-fit signal-to-noise ratio (SNR) $>3$ for all the four bands, no deblending required for profile-fitting and high fractional detection rate, given by ratio number frames on which a source was detected to the number of frames which were available for extraction (w?nm/w?m; where `?' represents different ALLWISE bands), close to unity. We visually inspected the ALLWISE Atlas images in all four bands for all the NYMG candidates.
 
The IR excess candidates are selected by constructing spectral energy distribution (SED) for all the stars. The SED displays flux density versus wavelength from the photometry. The SED for all the stars are constructed using the method given in \citet{shridharan2022emission} and fitted over BT-NextGen (AGSS2009) spectra model and mean sampled \textit{Gaia} BP/RP spectra (as shown in Figure \ref{fig:sed} for HD 37852). For candidates lacking adequate detection in the W3 and W4 bands, we examined the presence of alternative far-infrared data from missions like Spitzer and Herschel to perform SED fitting. From our sample of 3,153 stars, we detected 51 stars with IR excess (given in Table  \ref{tab: final tab}). Out of these 51, 19 stars are from the new NYMG candidates we identified in this work. HD 37852 in figure \ref{fig:sed} is one of the 19 new NYMG  candidates with IR excess. 

\begin{figure*}
    \centering
     \includegraphics[scale=0.5]{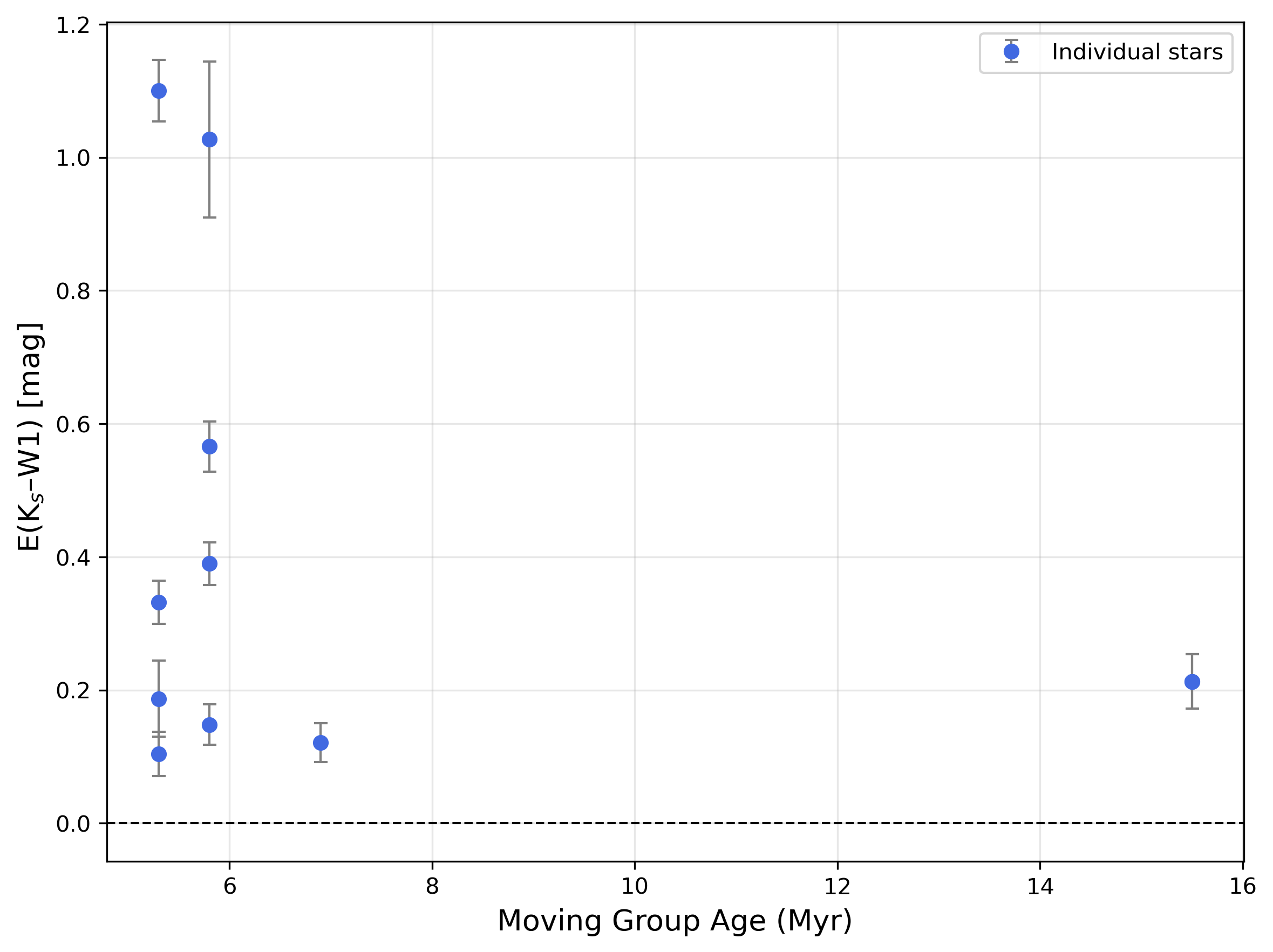}
    \caption{The relation between age and color excess in $E(K - W1)$. The blue points indicate the IR excess candidates from different NYMGs ages marked with the corresponding error bars. }
    \label{fig:excess_age}
\end{figure*}

The IR excess detections obtained in this study are consistent with those reported in previous works, and no new IR excess sources were identified. Figure~\ref{fig:excess_age} shows the variation of color excess $E(K - W1)$ as a function of age for the NYMG members. A clear decrease in the color excess with increasing age is evident, consistent with the expected disk dissipation over time. In this figure, we included only candidates exhibiting significant excesses ($>3\sigma$) in $K - W1$ and excluded those showing excess at longer wavelengths. The astrometric quality selection using the \textit{Gaia} RUWE parameter can influence the detection of near-IR excess, as inflated RUWE values may result in the exclusion of genuine excess candidates. However, a relaxed RUWE criterion would reduce the reliability of moving group membership assignments. As noted by \citet{roquette2025nemesis} the RUWE distributions between optically thick-disk and diskless samples show no statistically significant difference. A more rigorous analysis is therefore required to fully characterize the impact of these selection effects on IR excess detection. A detailed study regarding the properties of the IR excess candidates will be presented in a follow-up work.

\section{Summary} \label{discuss}
Most of the NYMG candidates are low-mass stars, for which complete kinematic information was not available for a long time in the literature. However, with the advent of the \textit{Gaia} space mission, we obtained photometric and astrometric information about such low-mass, younger stars in the galaxy. In the present study, we made use of the high quality astrometric and photometric data from \textit{Gaia} DR3, and, BANYAN $\Sigma$, one of the prominent algorithms widely used, for assigning membership probability. The major results from this study are summarized below;

\begin{enumerate}
    \item We provide a most recent and updated catalog of 3,153 candidates of 12 NYMGs within 150 pc, out of which 1651 are new NYMG candidates identified for the first time in this study. This fairly comprehensive catalog can be used as a reference for further in-depth study of NYMGs. From our new NYMG candidates, 29 candidates had the membership probability $>90\%$ in BANYAN-$\Sigma$ framework. We believe these 29 candidates will be a value addition to the previously known bonafide list of NYMGs given in ~\citet{gagne2018banyana} and \citet{gagne2018banyanc}.

    \item Previous studies have highlighted the need for refining the `Good-box’ criterion in the membership analysis of moving groups. We addressed this open question with a better and upgraded sample of NYMG to test the credibility of Good-box criterion. Space velocity components (U, V, W) values are calculated for 1,314 stars. Using these candidates, we proposed a new refined Good-box with the dimensions of U,V,W to be -29 to -3 km~s$^{-1}$, -30 to 2 km~s$^{-1}$, -18 to 2 km~s$^{-1}$, respectively. This criterion can be used as a preliminary step to segregate NYMG stars from a sample of the stellar population.

    \item In this study, we utilized precise astrometry data from \textit{Gaia} DR3 to compile an updated NYMG catalog and determine ages for highly probable members- both from literature and new candidates that have high membership probability- using high-quality photometry data from \textit{Gaia} DR3 and employing MIST isochrone models. Our analysis revealed a substantial scatter of NYMG members on the CMDs, resulting in the candidates from each NYMG intersecting with multiple MIST isochrones. Consequently, the isochrone ages presented in our study should be interpreted as indicative age ranges for the candidate members rather than absolute ages for the NYMGs themselves.

    \item Using an updated candidate list of 12 NYMGs from our study, we analyzed stars exhibiting IR excess. Due to the proximity of these moving groups and their (evolutionary favorable) age range, those stars with IR excess prove to be a very good set of candidates for further study on disk evolution and identification of new planetary systems. From our sample of 3,153 stars, 51 candidates are reported as candidates with IR excess, out of which 19 candidates are from the new NYMG candidates reported in this study. All of the IR excess candidates that we report in this study have been reported to have excess in the literature. Furthermore, we find that the excess color decreases with increasing age confirming the occuring disk dissipition over time. 

\end{enumerate}
\par
The significance of the current study is to provide a recent and upgraded list of NYMGs using precise astrometric data from \textit{Gaia} DR3. This catalog can serve as a dependable aid for further research in NYMGs. It is obvious that the further releases from \textit{Gaia} will help in identifying more candidates in the near future which can be an addition to the current list. We were able to address the open question of `Good-box' criterion for NYMGs with \textit{Gaia} data, that was previously discussed in literature. Our age determination analysis reveal that NYMG stars display a wide spread on the CMDs overlapping on several age isochrones. Furthermore, we also identified NYMG candidates that exhibit the presence of disks which consequently aids in exoplanet studies.

%%Use table* environment to get the table spanning both the columns

%\begin{table*}[htb]
%\tabularfont
%\caption{Caption text here}\label{secondTable}
%\begin{tabular}{lccccccccccccr}
%\topline
%\textbf{head1}&\multicolumn{11}{c}{\textbf{head2}}&\textbf{head3}\\
%\midline
%one& two &three&four&five&six&seven&eight&nine&ten&eleven&twelve&thirteen\\
%1&2&3&4&5&6&7&8&9&10&11&12&13\\
%aaa&bbbb&cccc&ddddd&eee&ffff&ggggg&hhhhhhhh&iiii&kkkkkk&hhh&jjjjjj&lllll\\
%\hline
%\end{tabular}
%\tablenotes{Table footnote here. Table spanning both the columns.}
%\end{table*}

%%An example of a figure

%\begin{figure}[!t]
%\includegraphics[width=.8\columnwidth]{fig1.eps}
%\caption{caption goes here}\label{figOne}
%\end{figure}

%%An example of a double column figure
%%Use figure* environment

%\begin{figure*}
%\centering\includegraphics[height=.15\textheight]{fig1.eps}
%\caption{caption spanning two columns}
%\centering\includegraphics[height=.25\textheight]{fig1.eps}
%\caption{caption here}
%\end{figure*}

\vspace{-2em}

\section*{Acknowledgements}

ML and UK thank Shridharan for his valuable suggestions throughout the course of the work. UK acknowledges the Department of Science and Technology (DST) for the INSPIRE FELLOWSHIP (IF180855). This work has made use of data from the European Space Agency
(ESA) mission \textit{Gaia} (https://www.cosmos.esa.int/Gaia), processed
by the \textit{Gaia} Data Processing and Analysis Consortium (DPAC;
https://www.cosmos.esa.int/web/Gaia/dpac/ consortium). Funding
for the DPAC has been provided by national institutions, in particular, the institutions participating in the \textit{Gaia} Multilateral Agreement.
Also, we made use of the VizieR catalog access tool, Simbad and
Aladdin, CDS, Strasbourg, France. 
\vspace{-1em}

%%use \balance somewhere in the left column of the last page to balance the two columns in the end page

\bibliography{ref4}
%%Appendix

\appendix

\section{Other space velocity dispersion criteria}\label{1appen}
As it is also described in the text above, the Good-box defined by \citet{zuckerman2004young}, is unable to incorporate the majority of the moving group candidates in this study. We thus introduce dispersion cuts of 1$\sigma$, 2$\sigma$ and 3$\sigma$ on the space velocities of the NYMG candidates identified in this study. The condition with a 2$\sigma$ dispersion cut was found to be best suited for our results. Other dispersion cuts like 1$\sigma$ and 3$\sigma$ are shown in the figures below where we observe that both the dispersion criteria are unsuitable for our sample of NYMGs.

\begin{figure} 
\includegraphics[width= \columnwidth]{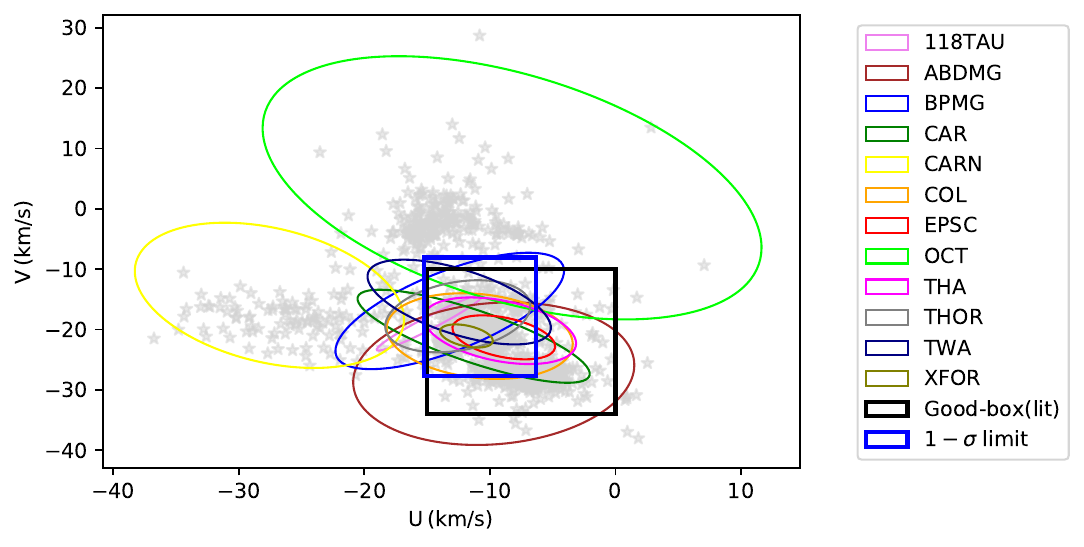}
\includegraphics[width= 0.75\columnwidth]{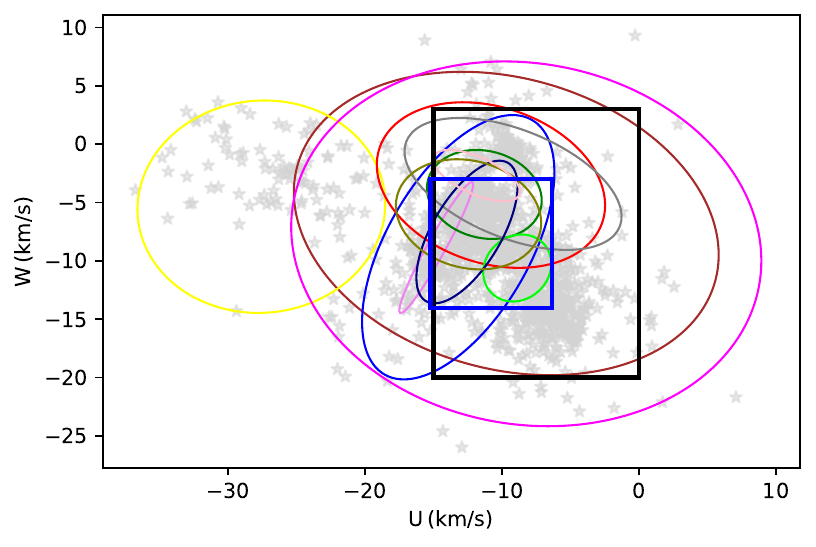}
\includegraphics[width= 0.75\columnwidth]{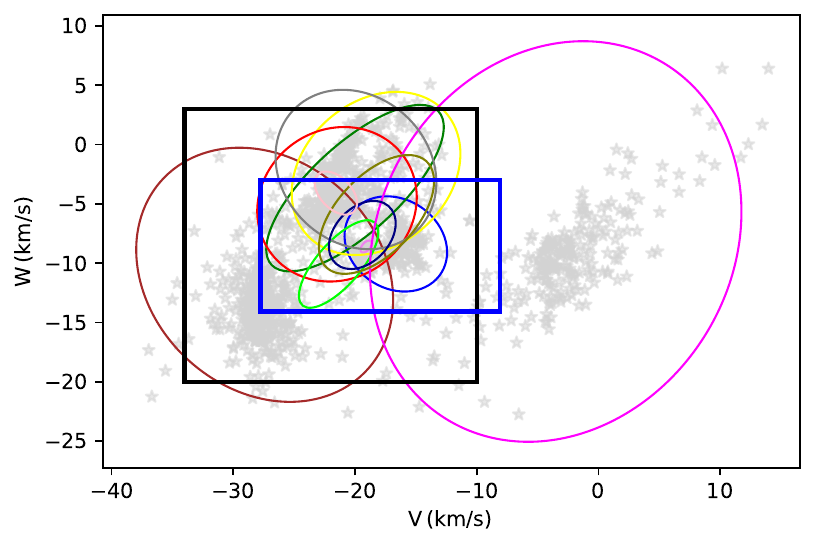}
\caption{Same as fig. \ref{sp gb}, with 1 $\sigma$ velocity dispersion limit over the Good-box.}
\label{1sigmaappen}
\end{figure}

\begin{figure}
\includegraphics[width= \columnwidth]{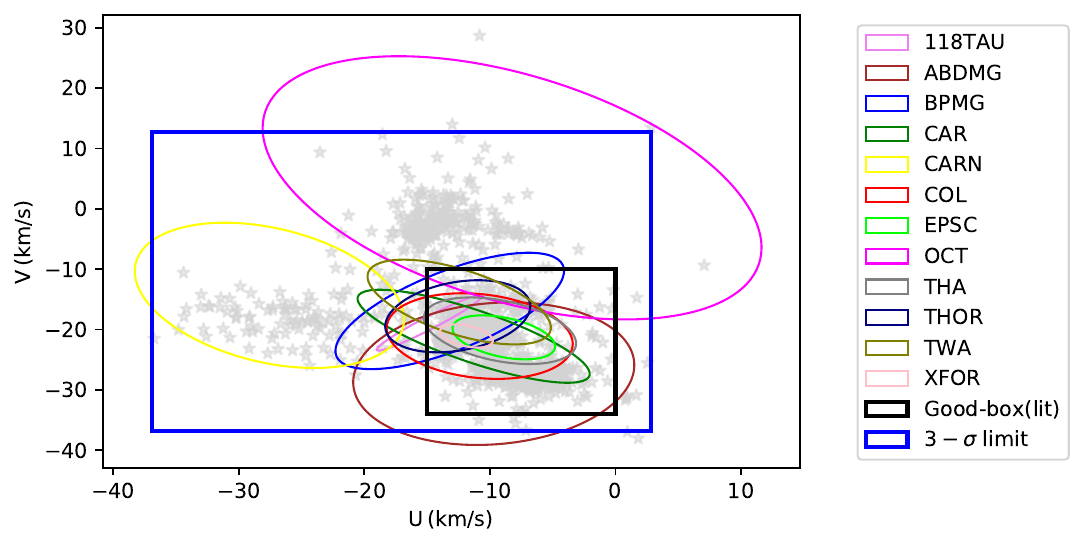}
\includegraphics[width= 0.75\columnwidth]{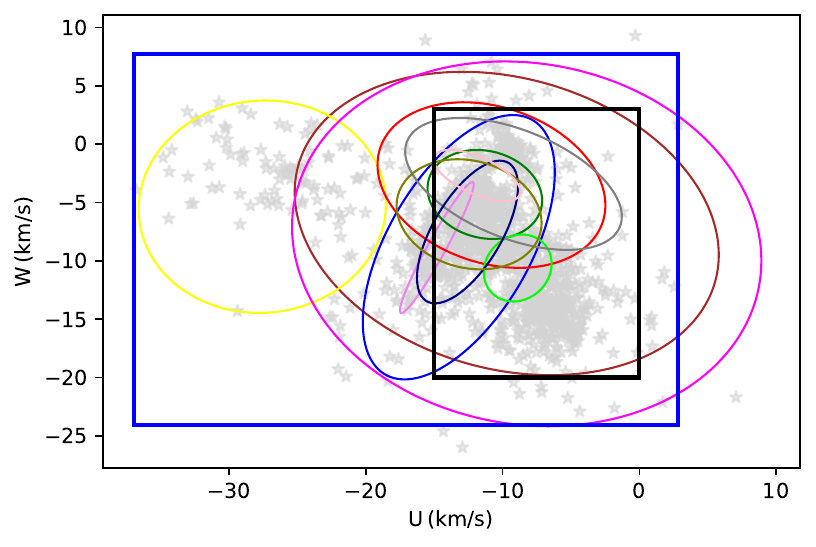}
\includegraphics[width= 0.75\columnwidth]{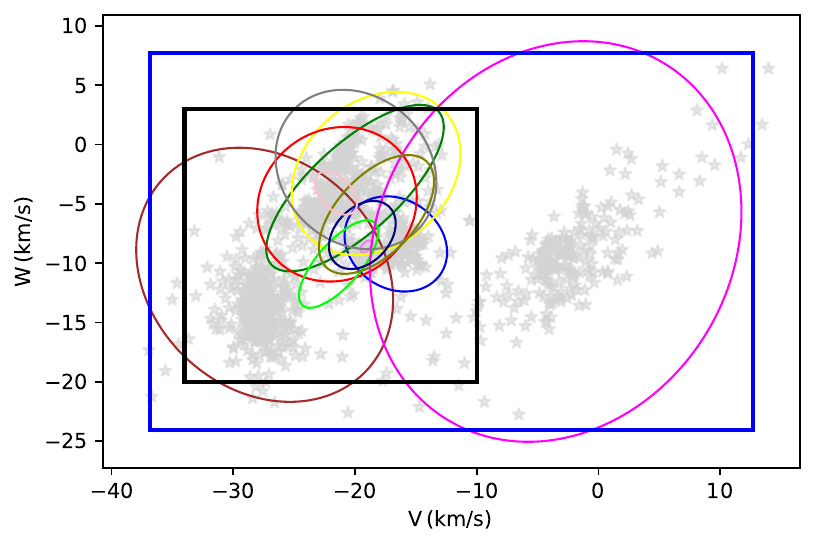}
\caption{Same as fig. \ref{sp gb}, with 1 $\sigma$ velocity dispersion limit over the Good-box.}
\label{3sigmaappen}
\end{figure}

\section{Total candidates within 150 pc from \textit{Gaia} DR3} \label{2appen}
\begin{figure}
\includegraphics[width= 1.5\columnwidth]{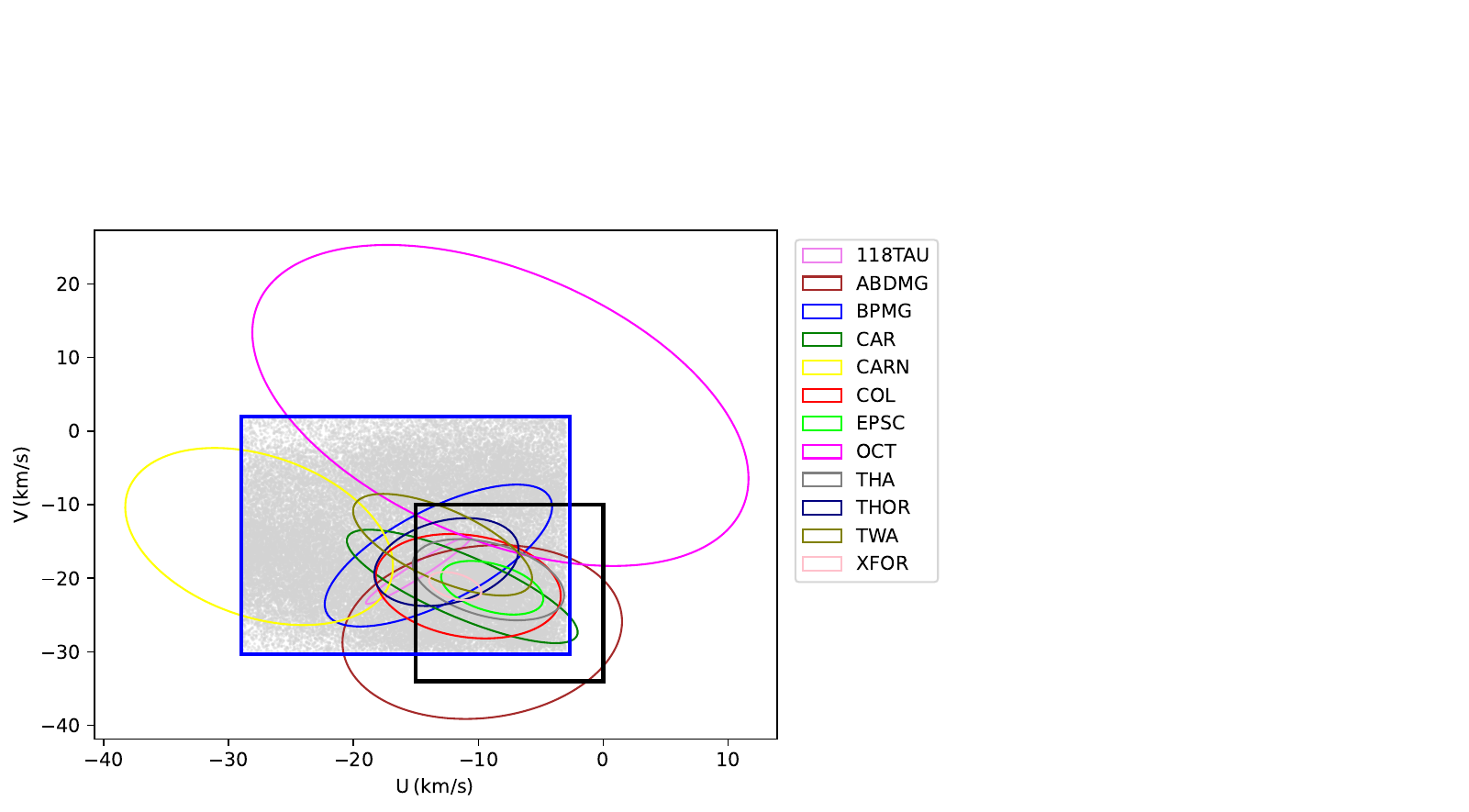}

\caption{Same as fig. \ref{sp gb} with the addition of all the \textit{Gaia} DR3 stars within 150 pc, marked in gray stars, that lie within our redefined Good-box.}
\label{app_150pc}
\end{figure}

As outlined in Section \ref{sample}, we applied a range of selection criteria to the sample of stars obtained from the \textit{Gaia} DR3 source catalog to minimize potential errors. These criteria included constraints on RUWE and proper motions, which significantly impact membership analysis. We additionally analyzed a scenario where we consider all the stars we obtained from \textit{Gaia} DR3 within 150 pc, visualizing their spatial velocities within the Good-box region, that we redefine in our study. The fig.\ref{app_150pc} illustrates 53,394 additional \textit{Gaia} candidates within 150 pc, excluding our final list of candidates, that fall within the Good-box limits. Constraining the errors in the astrometric parameters of these candidates can help us further classify them as members of any moving group or as field stars. Therefore, with future data releases from \textit{Gaia} providing improved astrometry for these stars, we have the potential to identify newer moving group members, thereby expanding the current list from this study.

%%Use section* for acknowledgements

\end{document}